\documentclass[11pt]{article}
\usepackage{amssymb,amsthm}

\title{Asymptotic form of two-point correlation function \\
  of the $XXZ$ spin chain}
\author{Yasuhiro Fujii\footnotemark[2]\ \ and Miki Wadati}
\date{\textit{\small Department of Physics, Graduate School of Science,
  University of Tokyo,\\  Hongo 7--3--1, Bunkyo-ku, Tokyo 113--0033, Japan}}

\newcommand{\rmi}{\mbox{i}}
\newcommand{\rme}{\mbox{e}}
\newcommand{\rmd}{\mbox{d}}
\newcommand{\rmii}{\mbox{\scriptsize i}}
\newcommand{\der}{\partial}
\renewcommand{\phi}{\varphi}
\newcommand{\bZ}{\mathbb{Z}}
\newcommand{\bC}{\mathbb{C}}

\theoremstyle{plain}
\newtheorem*{theorem}{Theorem}
\theoremstyle{remark}
\newtheorem{lemma}{Lemma}

\newcommand{\vac}{\mbox{vac}}
\newcommand{\2}{\mbox{$\frac{1}{2\pi}$}}
\newcommand{\Det}{\mbox{Det}}
\newcommand{\Tr}{\mbox{Tr}}
\newcommand{\tx}{\widetilde{\chi}}

\begin{document}
\maketitle
\footnotetext[2]{e-mail: \texttt{fujii@monet.phys.s.u-tokyo.ac.jp}}

\begin{abstract}
  Correlation functions of the $XXZ$ spin chain in the critical regime
  is studied at zero-temperature.
  They are exactly represented in the Fredholm determinant form
  and are related with an operator-valued Riemann--Hilbert problem.
  Analyzing this problem
  we prove that a two-point correlation function
  consisting of sufficiently separated spin operators
  is expressed by power-functions of the distance between those operators.
\end{abstract}

\section{Introduction}
\label{int}

We consider a two-point correlation function of the $XXZ$ spin chain
in an external magnetic field at zero-temperature.
The Hamiltonian is defined by
\begin{equation}
  H_{XXZ} =
  \sum_{n\in\bZ}(\sigma_n^x\sigma_{n+1}^x+\sigma_n^y\sigma_{n+1}^y
  +\Delta\sigma_n^z\sigma_{n+1}^z-h\sigma_n^z).
\end{equation}
The anisotropy parameter $\Delta$ is parameterized as
$\Delta=\cos 2\eta$ ($\frac{\pi}{2}<\eta<\pi$),
which implies the critical regime.
The Pauli matrices $\sigma_n^x$, $\sigma_n^y$, $\sigma_n^z$
act on the $n$-th site and $h$ indicates an external magnetic field.
The $XXZ$ spin chain is one of quantum solvable models
and is analyzed
by means of Quantum Inverse Scattering Method \cite{KBI}.
Based on this method 
the generating functional of correlation functions is
represented in the Fredholm determinant form \cite{EFIK}.

The determinant representation of the generating functional
has been investigated in some limiting cases.
In the limit of strong magnetic field $h\rightarrow h_c=4\cos^2\eta$
this is connected with the $\tau$-function of
the Painlev\'{e} $V$ equation \cite{EFItK1}.
On the basis of this equation
we can evaluate correlation functions of the $XXZ$ spin chain
in strong magnetic field \cite{FW1}.
The generating functional depends on
an extra parameter $\alpha$ (see (\ref{q}))
and in the limit $\alpha\rightarrow-\infty$
itself indicates a special correlation function
called the ferromagnetic-string-formation-probability (FSFP).
The FSFP describes the probability of finding
a ferromagnetic string of adjacent parallel spins
and the corresponding Fredholm determinant
is known to be related with
an operator-valued Riemann--Hilbert problem \cite{EFItK2}.
Thus the evaluation of FSFP is reduced to
the solution of this problem.
We remark that either the analysis of the Painlev\'{e} $V$ equation
or Riemann--Hilbert problem is a kind of
classical inverse scattering problems.
For special cases above,
the calculation of correlation functions of the $XXZ$ spin chain
can be reduced to a \textit{classical} inverse problem,
despite the model is \textit{quantum} mechanical.

Recently, the operator-valued Riemann--Hilbert problem
associated with the generating functional has been proposed \cite{FW2}.
Surprisingly, the calculation of \textit{any} correlation function
of the $XXZ$ spin chain is interpreted as
a classical inverse scattering problem.
In particular we are in a position to obtain
the closed forms of two-point correlation functions:\ 
there was no valid method to
evaluate two-point correlation functions of the $XXZ$ spin chain
from the standpoint of quantum solvable models,
because the integral representations of two-point correlation functions
obtained in \cite{JM} and \cite{KMT}
are too complicated to analyze them.
In this paper, on the basis of our Riemann--Hilbert problem,
we compute the large-$m$ asymptotic form of
a two-point correlation function $\langle\sigma_m^z\sigma_1^z\rangle$
and prove the following theorem:
\begin{theorem}
  For sufficiently large $m$
  a two-point correlation function
  $\langle\sigma_m^z\sigma_1^z\rangle$
  is expressed by power-functions of $m$.
\end{theorem}
This corresponds to a rigorous proof of the well-known conjecture that
two-point functions of the $XXZ$ spin chain
in the critical regime decay in power of
the distance between spin operators at zero-temperature,
which was derived by means of Conformal Field Theory,
the bosonization method, the numerical analysis and so on.
In order to prove the theorem,
in section~\ref{orhp},
we define the Fredholm determinant representation
of the generating functional
and formulate the associated operator-valued Riemann--Hilbert problem.
In section~\ref{ana} we analyze this problem
and determine the asymptotic form of generating functional.
As a result the theorem is proved.
The last section~\ref{conc} is devoted to concluding remarks.

\setcounter{equation}{0}
\section{Operator-valued Riemann--Hilbert problem}
\label{orhp}

In this paper we pay attention to a two-point correlation function
$\langle\sigma_m^z\sigma_1^z\rangle$,
which is computed through the generating functional.
In this section we define the Fredholm determinant representation
of the generating functional
and formulate the associated operator-valued Riemann--Hilbert problem.

\subsection{Fredholm determinant}

A two-point correlation function
$\langle\sigma_m^z\sigma_1^z\rangle$ is
given by the generating functional $Q^{(m)}(\alpha)$ via \cite{EFIK}
\begin{equation}
  \label{core}
  \langle\sigma_m^z\sigma_1^z\rangle =
  2\Delta_m\left.\frac{\der^2 Q^{(m)}(\alpha)}{\der\alpha^2}
  \right|_{\alpha=0}
  -4\left.\frac{\der Q^{(1)}(\alpha)}{\der\alpha}\right|_{\alpha=0}
  +1.
\end{equation}
Here $\Delta_m$ is the lattice Laplacian defined by
$\Delta_m f(m)=f(m)-2f(m-1)+f(m-2)$.
The generating functional $Q^{(m)}(\alpha)$ is represented by
the Fredholm determinant:
\begin{equation}
  \label{q}
  Q^{(m)}(\alpha) =
  Q_0\langle\vac|\Det(1-\2 V^{(m)})|\vac\rangle
\end{equation}
where $Q_0$ is a normalization factor.
The Fredholm determinant is expanded as
\begin{equation}
  \Det(1-\2 V^{(m)}) =
  \sum_{n=0}^\infty\frac{1}{n!}\left(-\frac{1}{2\pi}\right)^n
  \int_\Gamma\prod_{k=1}^n\rmd z_k\det{}_n V^{(m)}
\end{equation}
where $\det{}_n V^{(m)}$ is the normal determinant of
an $n\times n$ matrix $V^{(m)}(z_i,z_j)$ ($i,j=1,\ldots,n$).
The integration contour $\Gamma$
runs on a part of the unit circle anti-clockwise:
$\Gamma\ni\rme^{\rmii\theta}$ ($2\pi-\psi<\theta<\psi$).
The boundary value $\psi$ depends on
the anisotropy parameter $\eta$ and the external magnetic field $h$,
lies in $\pi<\psi<2\eta$
and approaches to $\pi$ in the strong magnetic field limit
$h\rightarrow h_c=4\cos^2\eta$,
to $2\eta$ in the limit $h\rightarrow 0$ \cite{KBI,FW2}.
The kernel $V^{(m)}$ is given by the scalar product
of four-component vectors as follows \cite{FW2}:
\begin{equation}
  V^{(m)}(z,w) =
  \rmi\int_0^\infty\frac{\rmd s}{z-w}
  \sum_{j=1}^4 a_j^{(m)}(z|s)b_j^{(m)}(w|s)
\end{equation}
where vectors are defined by
\begin{eqnarray}
  \mbox{\boldmath$a$}^{(m)}(z|s) &=&
  \left(\rmi(q-q^{-1})\frac{qz-1}{z-q}\right)^{1/2}
  \left(
    \begin{array}{c}
      z^{m/2}\exp(\rmi q^{-1}\frac{qz-1}{z-q}s+\alpha-\phi_3(z)) \\
      z^{m/2}\exp(-\rmi q\frac{qz-1}{z-q}s) \\
      z^{-m/2}\exp(\rmi q^{-1}\frac{qz-1}{z-q}s-\phi_2(z)) \\
      z^{-m/2}\exp(-\rmi q\frac{qz-1}{z-q}s+\alpha+\phi_1(z)-\phi_3(z))
    \end{array}
  \right)
  \nonumber \\\\
  \mbox{\boldmath$b$}^{(m)}(z|s) &=&
  \left(\rmi(q-q^{-1})\frac{qz-1}{z-q}\right)^{1/2}
  \left(
    \begin{array}{c}
      z^{-m/2}\exp(-\rmi q\frac{qz-1}{z-q}s+\phi_4(z)) \\
      -z^{-m/2}\exp(\rmi q^{-1}\frac{qz-1}{z-q}s) \\
      z^{m/2}\exp(-\rmi q\frac{qz-1}{z-q}s+\phi_2(z)) \\
      -z^{m/2}\exp(\rmi q^{-1}\frac{qz-1}{z-q}s-\phi_1(z)+\phi_4(z))
    \end{array}
  \right).
\end{eqnarray}
and $q=\rme^{2\rmii\eta}$.
The operators $\phi_j(z)$ ($j=1,\ldots,4$) are 
bosonic quantum fields called the \textit{dual fields}.
They are decomposed into coordinate and momentum fields as
$\phi_j(z)=q_j(z)+p_j(z)$ and act on the vacuum states defined by
$\langle\vac|q_j(z)=p_j(z)|\vac\rangle=0$
($j=1,\ldots,4$).
The commutation relation between coordinate and momentum fields
is expressed in the following matrix form:
\begin{equation}
  \label{comm}
  [q_j(z),p_k(w)] =
  \left(
    \begin{array}{cccc}
      1 & 0 & 1 & 0 \\
      0 & 1 & 0 & 1 \\
      0 & 1 & 1 & 1 \\
      1 & 0 & 1 & 1
    \end{array}
  \right)_{jk}\log h(z,w)
  +\left(
    \begin{array}{cccc}
      1 & 0 & 0 & 1 \\
      0 & 1 & 1 & 0 \\
      1 & 0 & 1 & 1 \\
      0 & 1 & 1 & 1
    \end{array}
  \right)_{jk}\log h(w,z)
\end{equation}
with
\begin{equation}
  h(z,w) =
  \frac{1}{q-q^{-1}}
  \left(
    q\sqrt{\frac{qz-1}{z-q}\frac{w-q}{qw-1}}
    -q^{-1}\sqrt{\frac{z-q}{qz-1}\frac{qw-1}{w-q}}
  \right).
\end{equation}
By definition the dual fields are commutative:
$[\phi_j(z),\phi_k(w)]=0$ ($j,k=1,\ldots,4$).
They are thus interpreted as entire functions
unless they act on the vacuum states
to produce the vacuum expectation values.
The following lemma is useful.
\begin{lemma}
  A dual field $\phi_4(z)$ is equal to $\phi_3(z)$.
\end{lemma}
\begin{proof}
  If a set of bosonic fields with four-species has
  the same commutation relation as (\ref{comm}),
  the kernel $V^{(m)}$ whose dual fields are
  replaced with those fields is equivalent to the original kernel.
  We put c-numbers $\alpha$ and $z^{\pm m/2}$ into the dual fields.
  Under a simultaneous replacement
  $\phi_1\leftrightarrow\phi_2$ and $\phi_3\leftrightarrow\phi_4$
  the commutation relation (\ref{comm}) is invariant
  and the kernel $V^{(m)}(z,w)$ is then changed into
  $\exp(\phi_3(w)-\phi_4(z))V^{(m)}(w,z)$.
  It thus follows that
  \begin{equation}
    V^{(m)}(z,z)=\exp(\phi_3(z)-\phi_4(z))V^{(m)}(z,z).
  \end{equation}
  Note that the kernel $V^{(m)}(z,z)$ contains the dual fields
  $\phi_1(z),\ldots,\phi_4(z)$.
  In order to produce the same vacuum expectation values
  in both sides,
  a dual field $\phi_4(z)$ must be equal to $\phi_3(z)$.
\end{proof}
Hearafter we replace a dual field $\phi_4(z)$ with $\phi_3(z)$
on the basis of this lemma.
By virtue of this lemma the recursion relation (\ref{pre}) contains
no dual fields and makes clear that two-point correlation functions
decay in power.

\subsection{Operator-valued Riemann--Hilbert problem}

The Fredholm determinant $\Det(1-\2 V^{(m)})$ is related with
\textit{matrix integral operators} \cite{FW2}.
They are defined by matrices whose multiplication requires
extra integral calculus:\ 
those products are formulated as
\begin{equation}
  \label{prod}
  (EF)_{ij}(z|s,t) =
  \int_0^\infty \rmd r\sum_k E_{ik}(z|s,r)F_{kj}(z|r,t).
\end{equation}
From now on we omit the dependency on extra variables $s$, $t$
of matrix integral operators unless necessary
and denote the delta-function $\delta(s-t)$ by $I$.
Consider a $4\times 4$ matrix integral operator $\chi^{(m)}(z)$
that satisfies the following conditions (a)--(c) \cite{FW2}:
\begin{enumerate}
\item[(a)] Each element of $\chi^{(m)}(z|s,t)$ is an analytic function
  of $z\in\bC\backslash\Gamma$ for any $m\in\bZ_{>0}$, $s,t\in\bC$.
\item[(b)] $\chi^{(m)}(\infty)=I$.
\item[(c)] Let $\chi_+^{(m)}(z)$ ($\chi_-^{(m)}(z)$) be
  the boundary value of $\chi^{(m)}(z)$
  on the left (the right) side of a positive direction
  of the oriented contour $\Gamma$.
  It then follows that
  \begin{equation}
    \label{xxl}
    \chi_-^{(m)}(z) = \chi_+^{(m)}(z)L^{(m)}(z).
    \qquad
    (z\in\Gamma)
  \end{equation}
\end{enumerate}
The $4\times 4$ matrix integral operator $L^{(m)}(z)$ is called
the \textit{conjugation matrix} and is expressed by
\begin{equation}
  L^{(m)}(z) =
  \left(
    \begin{array}{cccc}
      I-\kappa P(z) & \kappa\rme^{-\phi_3}Q(z) &
      -\kappa z^m\rme^{\phi_2-\phi_3}P(z) &
      \kappa z^m\rme^{-\phi_1}Q(z) \\
      -\rme^{\phi_3}R(z) & I+P^T(z) & -z^m\rme^{\phi_2}R(z) &
      z^m\rme^{-\phi_1+\phi_3}P^T(z) \\
      -z^{-m}\rme^{-\phi_2+\phi_3}P(z) &
      z^{-m}\rme^{-\phi_2}Q(z) &
      I-P(z) & \rme^{-\phi_1-\phi_2+\phi_3}Q(z) \\
      -\kappa z^{-m}\rme^{\phi_1}R(z) &
      \kappa z^{-m}\rme^{\phi_1-\phi_3}P^T(z) &
      -\kappa\rme^{\phi_1+\phi_2-\phi_3}R(z) & I+\kappa P^T(z)
    \end{array}
  \right)
\end{equation}
where $\kappa=\rme^\alpha$
and the $z$-dependency of the dual fields is omitted.
The integral operators $P(z)$, $Q(z)$, $R(z)$ and $P^T(z)$
are given below:
\begin{eqnarray}
  P(z) &=&
  \rmi(q-q^{-1})\frac{qz-1}{z-q}
  \exp\left(\rmi q^{-1}\frac{qz-1}{z-q}s
    -\rmi q\frac{qz-1}{z-q}t\right)
  \\
  Q(z) &=&
  \rmi(q-q^{-1})\frac{qz-1}{z-q}
  \exp\left(\rmi q^{-1}\frac{qz-1}{z-q}(s+t)\right)
  \\
  R(z) &=&
  \rmi(q-q^{-1})\frac{qz-1}{z-q}
  \exp\left(-\rmi q\frac{qz-1}{z-q}(s+t)\right)
  \\
  P^T(z) &=&
  \rmi(q-q^{-1})\frac{qz-1}{z-q}
  \exp\left(-\rmi q\frac{qz-1}{z-q}s
    +\rmi q^{-1}\frac{qz-1}{z-q}t\right).
\end{eqnarray}
They satisfy the following relations:
\begin{eqnarray}
  \label{pqr1}
  P(z)P(z) = P(z) &&
  P(z)Q(z) = Q(z) \\
  Q(z)R(z) = P(z) &&
  Q(z)P^T(z) = Q(z) \\
  R(z)P(z) = R(z) &&
  R(z)Q(z) = P^T(z) \\
  \label{pqr2}
  P^T(z)R(z) = R(z) &&
  P^T(z)P^T(z) = P^T(z).
\end{eqnarray}
Here these products implicitly contain
the integrals for extra variables $s$, $t$
like for (\ref{prod}).

To find the $4\times 4$ matrix integral operator
that satisfies (a)--(c) is referred as
the operator-valued Riemann--Hilbert problem.
By using its solution $\chi^{(m)}(z)$
the Fredholm determinant $\Det(1-\2 V^{(m)})$
is computed via the following lemma \cite{FW2}:
\begin{lemma}
  \label{lem}
  The Fredholm determinant $\Det(1-\2 V^{(m)})$ obeys
  \begin{equation}
    \frac{\Det(1-\2 V^{(m+1)})}{\Det(1-\2 V^{(m)})} =
    \Det(\gamma_2[\chi^{(m)}(0)]^{-1}\gamma_2)
  \end{equation}
  where $\gamma_2=\frac{1}{2}(1-\sigma^z\otimes 1)$.
\end{lemma}
\begin{proof}
  By Lemma 5 and (4.7) of \cite{FW2} we have
  \begin{equation}
    \frac{\Det(1-\2 V^{(m+1)})}{\Det(1-\2 V^{(m)})} =
    \exp\Tr\log(\gamma_1+\gamma_2[\chi^{(m)}(0)]^{-1}\gamma_2)
  \end{equation}
  where $\gamma_1=\frac{1}{2}(1+\sigma^z\otimes 1)$.
  Due to the orthogonality $\gamma_1\gamma_2=0$ and
  $\gamma_1^2=\gamma_1$, $\gamma_2^2=\gamma_2$ it follows that
  \begin{eqnarray}
    \exp\Tr\log(\gamma_1+\gamma_2[\chi^{(m)}(0)]^{-1}\gamma_2) &=&
    \exp\Tr(\log\gamma_1+\log(\gamma_2[\chi^{(m)}(0)]^{-1}\gamma_2))
    \nonumber \\ &=&
    \exp\Tr\log(\gamma_2[\chi^{(m)}(0)]^{-1}\gamma_2).
  \end{eqnarray}
  Using the relation $\log\Det=\Tr\log$
  we obtain the lemma.
\end{proof}
On the basis of this lemma
we derive the large-$m$ asymptotic representation
of a two-point correlation function $\langle\sigma_m^z\sigma_1^z\rangle$
from the solution $\chi^{(m)}(z)$ of
operator-valued Riemann--Hilbert problem (a)--(c).

\setcounter{equation}{0}
\section{Asymptotic form of two-point correlation function}
\label{ana}

In this section we compute the large-$m$ asymptotic behavior
of the solution $\chi^{(m)}(z)$
of operator-valued Riemann--Hilbert problem (a)--(c).
By virtue of the relations (\ref{pqr1})--(\ref{pqr2})
the conjugation matrix $L^{(m)}(z)$ can be decomposed into
the product of triangular matrix integral operators as follows:
\begin{equation}
  \label{decom}
  L^{(m)}(z) = M^{(m)}(z)N^{(m)}(z)
\end{equation}
where
\begin{equation}
  M^{(m)}(z) =
  \left(
    \begin{array}{cccc}
      I & \frac{\kappa}{1+\kappa}\rme^{-\phi_3}Q(z) &
      -z^m\rme^{\phi_2-\phi_3}P(z) &
      \frac{\kappa}{1+\kappa}z^m\rme^{-\phi_1}Q(z) \\
      0 & I & -\frac{1}{\kappa}z^m\rme^{\phi_2}R(z) &
      \frac{1}{1+\kappa}z^m\rme^{-\phi_1+\phi_3}P^T(z) \\
      0 & 0 & I &
      \frac{1}{1+\kappa}\rme^{-\phi_1-\phi_2+\phi_3}Q(z) \\
      0 & 0 & 0 & I
    \end{array}
  \right)
\end{equation}
and
\begin{equation}
  N^{(m)}(z) =
  \left(
    \begin{array}{cccc}
      I-\frac{\kappa}{1+\kappa}P(z) & 0 & 0 & 0 \\
      -\frac{1}{\kappa}\rme^{\phi_3}R(z) &
      I+\frac{1}{\kappa}P^T(z) & 0 & 0 \\
      -\frac{1}{1+\kappa}z^{-m}\rme^{-\phi_2+\phi_3}P(z) &
      \frac{1}{1+\kappa}z^{-m}\rme^{-\phi_2}Q(z) &
      I-\frac{1}{1+\kappa}P(z) & 0 \\
      -\kappa z^{-m}\rme^{\phi_1}R(z) &
      \kappa z^{-m}\rme^{\phi_1-\phi_3}P^T(z) &
      -\kappa\rme^{\phi_1+\phi_2-\phi_3}R(z) & I+\kappa P^T(z)
    \end{array}
  \right).
\end{equation}
Let $\Gamma_1$ ($\Gamma_2$) be a contour
that runs from $\rme^{-\rmii\psi}$ to $\rme^{\rmii\psi}$
inside (outside) the unit circle.
Consider a matrix integral operator $\tx^{(m)}(z)$ defined below,
like for the case of the lattice sine--Gordon model \cite{EFItK3}.
\begin{enumerate}
\item[(i)] $\tx^{(m)}(z)=\chi^{(m)}(z)$
  outside the region enclosed by the contours
  $\Gamma_1$ and $\Gamma_2$.
\item[(ii)] $\tx^{(m)}(z) = \chi^{(m)}(z)M^{(m)}(z)$
  in the region enclosed by $\Gamma$ and $\Gamma_1$.
\item[(iii)] $\tx^{(m)}(z) = \chi^{(m)}(z)(N^{(m)}(z))^{-1}$
  in the region enclosed by $\Gamma$ and $\Gamma_2$.
\end{enumerate}
By definition $\tx^{(m)}(z)$ is analytic for
$z\in\bC\backslash(\Gamma_1\cup\Gamma_2)$ and obeys
\begin{eqnarray}
  \tx_-^{(m)}(z) &=& \tx_+^{(m)}(z)M^{(m)}(z)
  \qquad
  (z\in\Gamma_1)
  \\
  \tx_-^{(m)}(z) &=& \tx_+^{(m)}(z)N^{(m)}(z).
  \qquad
  (z\in\Gamma_2)
\end{eqnarray}
We are interested in the asymptotic behavior of
the solution $\chi^{(m)}(z)$ for sufficiently large $m$.
Notice that in the limit $m\rightarrow\infty$
the conjugation matrices $M^{(m)}(z)$ and $N^{(m)}(z)$
become block-diagonal
in the vicinities of contours $\Gamma_1$ and $\Gamma_2$,
respectively.
Take this limit and consider the original conjugation matrix
for $\chi^{(m)}(z)$ again.
Then we have
\begin{equation}
  \chi^{(\infty)}(z) =
  \left(
    \begin{array}{cc}
      \Phi_1(z) & 0 \\
      0 & \Phi_2(z)
    \end{array}
  \right).
\end{equation}
The $2\times 2$ matrix integral operators
$\Phi_1(z)$, $\Phi_2(z)$ satisfy
\begin{equation}
  \label{fgf}
  \Phi_a^-(z) = \Phi_a^+(z)G_a(z)
  \qquad
  (z\in \Gamma, \quad a=1,2)
\end{equation}
and the corresponding conjugation matrices are expressed by
\begin{eqnarray}
  G_1(z) &=&
  \left(
    \begin{array}{cc}
      I-P(z) & \rme^{-\phi_3}Q(z) \\
      -\frac{1}{\kappa}\rme^{\phi_3}R(z) & I+\frac{1}{\kappa}P^T(z)
    \end{array}
  \right)
  \\
  G_2(z) &=&
  \left(
    \begin{array}{cc}
      I-P(z) & \rme^{-\phi_1-\phi_2+\phi_3}Q(z) \\
      -\kappa\rme^{\phi_1+\phi_2-\phi_3}R(z) &
      I+\kappa P^T(z)
    \end{array}
  \right).
\end{eqnarray}
The calculation of a two-point correlation function
$\langle\sigma_m^z\sigma_1^z\rangle$ requires
the Fredholm determinant of the solution $\chi^{(m)}(z)$.
Taking the Fredholm determinants in both sides of (\ref{fgf})
we obtain the scalar Riemann--Hilbert problems:
\begin{equation}
  \label{phi}
  \Det\Phi_a^-(z) = \Det\Phi_a^+(z)\Det G_a(z).
  \qquad
  (z\in\Gamma,\quad a=1,2)
\end{equation}
In the same way as \cite{EFItK3}
the Fredholm determinants of $G_1(z)$, $G_2(z)$ are computed as
\begin{equation}
  \Det G_1(z) = \kappa^{-1}
  \qquad
  \Det G_2(z) = \kappa.
\end{equation}
According to the condition at the infinity:
$\Det\Phi_1(\infty)=\Det\Phi_2(\infty)=1$,
the solutions of scalar problems (\ref{phi})
are determined as follows:
\begin{equation}
  \label{sol}
  \Det\Phi_1(z) =
  \left(\frac{z-\rme^{-\rmii\psi}}{z-\rme^{\rmii\psi}}
  \right)^{\rmii\alpha/2\pi}
  \qquad
  \Det\Phi_2(z) =
  \left(\frac{z-\rme^{\rmii\psi}}{z-\rme^{-\rmii\psi}}
  \right)^{\rmii\alpha/2\pi}.
\end{equation}
They are analytic for $z\in\bC\backslash\Gamma$.

We evaluate the $m$-dependency of the solution $\chi^{(m)}(z)$.
Based on the decomposition (\ref{decom})
the conjugation matrix for sufficiently large $m$
is expressed as follows:
\begin{equation}
  L^{(m)}(z) =
  \left(
    \begin{array}{cc}
      G_1(z) & z^m G_{12}(z) \\
      z^{-m}G_{21}(z) & G_2(z)
    \end{array}
  \right)
\end{equation}
where $G_{12}(z)$, $G_{21}(z)$ are
$2\times 2$ matrix integral operators independent of $m$.
Let the solution $\chi^{(m)}(z)$ denote by
\begin{equation}
  \chi^{(m)}(z) =
  \left(
    \begin{array}{cc}
      X_1^{(m)}(z) & X_{12}^{(m)}(z) \\
      X_{21}^{(m)}(z) & X_2^{(m)}(z)
    \end{array}
  \right).
\end{equation}
Here $X_{12}^{(m)}(z)$, $X_{21}^{(m)}(z)$ decay
sufficiently rapidly as $m\rightarrow\infty$.
By the relation (\ref{xxl}) of the original Riemann--Hilbert problem,
the $2\times 2$ matrix integral operators
$X_1^{(m)}(z)$, $X_2^{(m)}(z)$ obey
\begin{equation}
  \begin{array}{rll}
    [X_1^{(m)}(z)]_- &=&
    [X_1^{(m)}(z)]_+ G_1(z)+z^{-m}[X_{12}^{(m)}(z)]_+ G_{21}(z) \\\
    [X_2^{(m)}(z)]_- &=&
    [X_2^{(m)}(z)]_+ G_2(z)+z^m[X_{21}^{(m)}(z)]_+ G_{12}(z).
  \end{array}
  \qquad
  (z\in\Gamma)
\end{equation}
In terms of the relation (\ref{fgf})
the conjugation matrix
$G_a(z)$ is given by $[\Phi_a^+(z)]^{-1}\Phi_a^-(z)$ ($a=1,2$).
Applying them we have
\begin{eqnarray}
  [X_1^{(m)}(z)]_-[\Phi_1^-(z)]^{-1} &=&
  [X_1^{(m)}(z)]_+[\Phi_1^+(z)]^{-1}
  +z^{-m}[X_{12}^{(m)}(z)]_+ G_{21}(z)[\Phi_1^-(z)]^{-1} \\\
  [X_2^{(m)}(z)]_-[\Phi_2^-(z)]^{-1} &=&
  [X_2^{(m)}(z)]_+[\Phi_2^+(z)]^{-1}
  +z^m[X_{21}^{(m)}(z)]_+ G_{12}(z)[\Phi_2^-(z)]^{-1}.
\end{eqnarray}
Here the variable $z$ lies on the contour $\Gamma$.
We remark that the inverse of $\Phi_a(z)$ exists
because $\Det\Phi_a(z)$ is non-zero for $z\in\Gamma$ ($a=1,2$)
(see (\ref{sol})).
By means of the Plemelj formulae (see \cite{M})
and the condition $X_1^{(m)}(\infty)=X_2^{(m)}(\infty)=I$,
the $2\times 2$ matrix integral operators
$X_1^{(m)}$, $X_2^{(m)}(z)$ are expressed below:
\begin{eqnarray}
  \label{x1}
  X_1^{(m)}(z) &=&
  \left(I
    -\frac{1}{2\pi\rmi}\int_\Gamma\frac{w^{-m}\rmd w}{w-z}
    [X_{12}^{(m)}(w)]_+ G_{21}(w)[\Phi_1^-(w)]^{-1}\right)\Phi_1(z) \\
  \label{x2}
  X_2^{(m)}(z) &=&
  \left(I
    -\frac{1}{2\pi\rmi}\int_\Gamma\frac{w^m\rmd w}{w-z}
    [X_{21}^{(m)}(w)]_+ G_{12}(w)[\Phi_2^-(w)]^{-1}\right)\Phi_2(z).
\end{eqnarray}
A two-point correlation function $\langle\sigma_m^z\sigma_1^z\rangle$
is computed by using only $[\chi^{(m)}(0)]^{-1}$.
In the case $z=0$
the right sides yield some terms whose order is less than $1/m$,
because the contour $\Gamma$ is a part of the unit circle
and $X_{12}^{(m)}(z)$, $X_{21}^{(m)}(z)$ decay
sufficiently rapidly as $m\rightarrow\infty$.
Hence we obtain the large-$m$ asymptotic behavior
of the solution $\chi^{(m)}(0)$ as follows:
\begin{equation}
  \label{x}
  \chi^{(m)}(0) =
  \left(
    \begin{array}{cc}
      \Phi_1(0)(I+O(m^{-1})) & 0 \\
      0 & \Phi_2(0)(I+O(m^{-1}))
    \end{array}
  \right).
  \qquad
  (m\gg 1)
\end{equation}

We are in a position to compute a two-point correlation function
$\langle\sigma_m^z\sigma_1^z\rangle$.
Applying the expressions (\ref{sol}) and (\ref{x})
to Lemma~\ref{lem} we have
\begin{equation}
  \label{pre}
  \frac{\Det(1-\2 V^{(m+1)})}{\Det(1-\2 V^{(m)})} =
  \exp\left(\frac{\psi}{\pi}\alpha\right)(1+O(m^{-1})).
  \qquad
  (m\gg 1)
\end{equation}
Since the leading term $\exp(\psi\alpha/\pi)$
does not contain any dual field,
the vacuum expectation value of the left hand side
can be written as
\begin{equation}
  \frac{Q^{(m+1)}(\alpha)}{Q^{(m)}(\alpha)} =
  \exp\left(\frac{\psi}{\pi}\alpha\right)(1+O(m^{-1})).
  \qquad
  (m\gg 1)
\end{equation}
From this recursion relation
we derive the large-$m$ asymptotic form of the generating functional
as follows:
\begin{equation}
  \label{Q}
  Q^{(m)}(\alpha) =
  f(m,\alpha)\exp\left(\frac{\psi}{\pi}\alpha m\right).
  \qquad
  (m\gg 1)
\end{equation}
Here $f(m,\alpha)$ satisfies
$f(m+1,\alpha)/f(m,\alpha)=1+O(m^{-1})$.
Substituting this asymptotic form into the relation (\ref{core})
we arrive at the following relation
for two-point correlation functions:
\begin{equation}
  \frac{\langle\sigma_{m+1}^z\sigma_1^z\rangle}
  {\langle\sigma_m^z\sigma_1^z\rangle} =
  1+O(m^{-1}).
  \qquad
  (m\gg 1)
\end{equation}
Thus the asymptotic form of a two-point correlation function 
of the $XXZ$ spin chain is shown to be given
by power-functions of the distance between spin operators.
This result is nothing but the theorem in section~\ref{int}.

\section{Concluding remarks}
\label{conc}

We have analyzed the operator-valued Riemann--Hilbert problem
associated with the generating functional of correlation functions
and have obtained the large-$m$ asymptotic form
of a two-point correlation function $\langle\sigma_m^z\sigma_1^z\rangle$
of the $XXZ$ spin chain.
This correlation functions is exactly expressed by power-functions.
Our approach based on Riemann--Hilbert problems
is useful for the analysis of long distance correlations.
In a forthcoming publication we will determine the exponent of power,
calculating the integrals (\ref{x1}) and (\ref{x2}).
The exponent is expected to depend on the anisotropy parameter $\eta$
and the external magnetic field $h$.

As stated in section~\ref{int},
a special correlation function FSFP
is obtained from the generating functional
by taking the limit $\alpha\rightarrow-\infty$.
However, in this limit,
our asymptotic expression (\ref{Q}) loses its meaning.
Similar situation arises from the asymptotic analysis
of FSFP of the $XXX$ spin chain ($\Delta=1$) \cite{FIK}.
From the viewpoint of Riemann--Hilbert problem
the evaluation of FSFP is more difficult
than that of the generating functional.
It is an open problem to compute the asymptotic form of
FSFP for the $XXX$ and $XXZ$ spin chain.
It is worth mentioning that
in the free fermion point $\Delta=0$
the corresponding FSFP has been shown to decay in Gaussian form \cite{DIZ}.

In this paper we have deformed
an operator-valued Riemann--Hilbert problem
for $4\times 4$ matrix integral operators to scalar problems,
which has been studied in detail \cite{M}.
In general,
it is hard to solve operator-valued Riemann--Hilbert problem:\ 
even in the case that the conjugation matrix has no integral operator
the solution can not be found in the closed form,
and for operator-valued case
its existence and uniqueness have not yet been proved.
They are fundamental and unsolved questions.
The investigation of operator-valued Riemann--Hilbert problem
is important
since it contains many applications to mathematics and physics.

\section*{Acknowledgment}

The authors wish to thank Professor T. Deguchi and Dr T. Tsuchida
for useful comments.


\begin{thebibliography}{99}
\bibitem{KBI} V. E. Korepin, N. M. Bogoliubov, A. G. Izergin:
  ``\textit{Quantum Inverse Scattering Method and Correlation Functions}'',
  (Cambridge University Press, 1993, Cambridge).
\bibitem{EFIK} F. H. L. E\ss ler, H. Frahm, A. G. Izergin
  and V. E. Korepin:
  \textit{Commun. Math. Phys.} \textbf{174} (1995) 191--214.
\bibitem{EFItK1} F. H. L. E\ss ler, H. Frahm, A. R. Its
  and V. E. Korepin:
  \textit{J. Phys.} A \textbf{29} (1996) 5619--5626.
\bibitem{FW1} Y. Fujii and M. Wadati:
  \textit{J. Phys. Soc. Jpn} \textbf{68} (1999) 3227--3235.
\bibitem{EFItK2} F. H. L. E\ss ler, H. Frahm, A. R. Its
  and V. E. Korepin:
  \textit{Nucl. Phys.} B \textbf{446} (1995) 448--460.
\bibitem{FW2} Y. Fujii and M. Wadati:
  \textit{J. Phys.} A \textbf{33} (2000) 1351--1361.
\bibitem{JM} M. Jimbo M and T. Miwa:
  ``\textit{Algebraic Analysis of Solvable Lattice Models}'',
  (CBMS Regional Conference Series in Mathematics \textbf{85}, 1995, AMS).
\bibitem{KMT} N. Kitanine, J. M. Maillet and V. Terras:
  \textit{Nucl. Phys.} B \textbf{567} (2000) 554--582.
\bibitem{EFItK3} F. H. L. E\ss ler, H. Frahm, A. R. Its
  and V. E. Korepin:
  \textit{J. Phys.} A \textbf{30} (1997) 219--244.
\bibitem{M} N. I. Muskhelishvili:
  ``\textit{Singular Integral Equations: Boundary Problems
    of Function Theory and Their Applications to Mathematical Physics}'',
  (Second edition, Dover Publications, 1992, New York).
\bibitem{FIK} H. Frahm, A. R. Its and V. E. Korepin:
  ``\textit{An operator-valued Riemann--Hilbert problem associated
    with the $XXX$ model}'',
  (CRM Proc. Lecture Notes \textbf{9}, AMS, 1996) 133--142.
\bibitem{DIZ} P. Deift, A. R. Its and X. Zhou:
  \textit{Ann. Math.} \textbf{146} (1997) 149--235.
\end{thebibliography}
\end{document}